# Thermally driven outflows from pair-plasma pressure-mediated shock surfaces around Schwarzschild black holes

Tapas K. Das★†
*Theoretical Astrophysics Group, S. N. Bose National Centre for Basic Sciences, Block JD, Sector III, Salt Lake, Calcutta 700 091, India*

**ABSTRACT**

Introducing a spherical, steady, self-supported pair-plasma pressure-mediated shock surface around a Schwarzschild black hole as the effective physical atmosphere that may be responsible for the generation of astrophysical mass outflows from relativistic quasi-spherical accretion, we calculate the mass outflow rate $R_{\dot{m}}$ by simultaneously solving the set of equations governing transonic polytropic accretion and isothermal winds. $R_{\dot{m}}$ is computed in terms of *only three* inflow parameters, which, we believe, has been done for the first time in our work. We then study the dependence of $R_{\dot{m}}$ on various inflow as well as shock parameters, and establish the fact that the outflow rate is essentially controlled by the post-shock proton temperature.

## 1 INTRODUCTION

In our previous works (Das 1999a, hereafter D99a; Das 1999b) we modelled the origin of the astrophysical mass outflow from matter accreting quasi-spherically on to Schwarzschild black holes. It has been shown (D99a) that it is possible to compute mass outflow rate (the measure of the fraction of accreting matter being blown away as wind) $R_{\dot{m}}$ from the vicinity of the black hole by self-consistent combination of exact transonic accretion–wind topologies. Taking the pair-plasma pressure-mediated shock [a standing, collisionless, self-supported shock proposed by Protheroe & Kazanas (1983, hereafter PK83) and Kazanas & Ellison (1986, hereafter KE86)] as the effective physical barrier around the black hole, the absolute value of $R_{\dot{m}}$ was computed in terms of the inflow parameters, and the dependence of $R_{\dot{m}}$ on those parameters was studied. The accreting matter as well as the outflow were assumed to obey the polytropic equation of state in the model proposed in D99a, while thermally driven outflow was not analysed. In this work we have dealt with isothermal outflow to demonstrate explicitly the temperature dependence of $R_{\dot{m}}$. Thus thermally driven outflows have been taken care of.

Owing to the fact that, close to a spherically or almost spherically accreting Schwarzschild black hole, the electron number density $n_e$ falls off as $r^{-3/2}$ while the photon number density $n_\gamma$ falls off as $r^{-2}$ (Frank, King & Raine 1995; see also McCray 1979), the ratio of $n_e$ to $n_\gamma$ is proportional to $\sqrt{r}$. This tells us that, with decrease of radial distance measured from the hole, a lesser number of electrons would be available per photon. Thus we may conclude that close to a quasi-spherically accreting black hole, where the shock-generated 'virtual surface' may form, momentum transfer by photons on the slowly outgoing and expanding subsonic outflow might have been a quite efficient process. Hence taking the outflow to be isothermal (at least up to the sonic point) would not be an unjustified assumption owing to the fact that momentum deposition of photons on to the outflow helps to keep the outflow temperature roughly constant as in the case of ordinary stellar winds (Tarafdar 1988 and references therein). Thus we take the isothermal equation of state for the outflow in this work to take care of thermally driven winds from matter accreting quasi-spherically with almost zero intrinsic angular momentum; whereas for the inflow part, we deal with polytropic accretion as in D99a. So, in some sense, this work represents some sort of 'hybrid model', which successfully connects flows obeying two different equations of state.

Our present work, we believe, has some important consequences regarding modelling the astrophysical outflows emanating from Galactic and extragalactic sources powered by isolated accreting compact objects. Galactic and extragalactic sources of jets and outflows are now widely believed to be fed by accreting black holes sitting at the dynamical centre of these sources. In the absence of any binary companion, quasi-spherical Bondi (1952) type accretion may occur on to an isolated central black hole if the accreting matter has a negligible amount of intrinsic angular momentum. On the other hand, unlike ordinary stellar bodies, black holes do not have their own 'physical' atmosphere, and outflows in these cases have to be generated from the accreting matter only. Hence we believe that it is *necessary* to study the accretion and outflow (from various astrophysical sources powered by accreting compact objects) in the same framework instead of treating the outflow separately from the accretion phenomenon. At the same time, as the fundamental criterion for constructing any self-consistent physical model demands the minimization of the number of inputs to the model, we may conclude that modelling the astrophysical outflow needs a concrete formulation where the outflow can be described in terms of the *minimum* number of physical parameters governing the inflow. The success of our work presented here, we believe, is precisely this. We could rigorously compute the mass outflow rate $R_{\dot{m}}$ in terms of *only three* inflow parameters, namely the specific energy of the inflow $\mathcal{E}$, the mass accretion rate $\dot{M}_{\rm in}$ (measured in units of the Eddington rate $\dot{M}_{\rm Edd}$) and the polytropic index $\gamma$ of the *inflow*. This three-parameter input set $\{\mathcal{E}, \dot{M}_{\rm in}, \gamma\}$ will be referred to as $\{P_3\}$ from now on throughout this paper.

★ Present address: Inter University Centre for Astronomy and Astrophysics (IUCAA), Post Bag 4, Ganeshkhind, Pune 411 007, India.
† tapas@iucaa.ernet.in

Not only do we provide a sufficiently plausible estimation of $R_{\dot{m}}$ in terms of only these three inflow parameters, we also successfully study the dependence and variation of this rate on various physical parameters governing the inflow. To the best of our knowledge, there is no such model available in the literature that rigorously studies the astrophysical outflow *only* in terms of the above-mentioned three parameters as has been done in this work.

Before going into the detail, a general outline of the flow characteristics in our model is presented below. We consider low-energy quasi-spherical accretion on to a Schwarzschild black hole. In KE86 it has been shown that for such accretion the kinetic energy of the infalling material may be randomized by proposing a steady, collisionless, self-supported, relativistic pair-plasma pressure-mediated shock. Relativistic protons are produced, which are shock-accelerated via first-order Fermi acceleration (Axford, Leer & Skadron 1977; Bell 1978a,b; Blandford & Ostriker 1978; Axford 1981a,b; Cowsik & Lee 1981; Blandford & Eichler 1987; Kirk & Duffy 1999). Because of the fact that the kinetic energy of these energetic protons is much higher than their gravitational potential energy (PK83), they are not readily captured by the black hole and are able to provide outward pressure sufficient to support the shock (for details see KE86 and references therein). Thus a spherical, steady and self-supported shock surface around the black hole is obtained. As the condition necessary for the development and maintenance of such a shock is satisfied when the Mach number of the *inflow* at shock stand-off distance ($M_{\rm sh}$) is considerably high (Ellison & Eichler 1984), we concentrate on polytropic accretion for our model, which produces a high shock Mach number solution.

We compute the shock location $r_{\rm sh}$ in terms of our three-parameter input set $\{P_3\}$. At the shock surface, the density of matter shoots up and inflow velocity falls off abruptly. In other words, highly supersonic inflow becomes subsonic and accreting matter becomes shock-compressed at this surface. Matter starts getting hotter and denser, and starts piling up on the surface. This hot and dense matter then gets a 'kick' due to the pair-plasma pressure and thermal pressure (the pressure generated by the high temperature produced at the shock), and a fraction of the accreting matter comes out as outflow. Though the pair-plasma pressure plays a crucial role in maintaining the shock and contributing the outward thrust for generating the outflow, as we will show in the next sections, the outflow is essentially thermally driven, and the main contribution to the generation of the outflow comes from the post-shock proton temperature, which comes out to be considerably high for our model. Thus a hot and dense spherical shock surface serves as the 'effective' physical atmosphere regarding the generation of mass outflow from matter accreting on to black holes. We calculate the mass outflow rate $R_{\dot{m}}$ from this shock surface by simultaneously solving the equations governing an exact transonic inflow–outflow system, and we study the dependence of this rate on various physical parameters governing the *inflow*.

It is to be mentioned here that one fundamental criterion for the formation of hydrodynamic outflows is that the outflowing wind should have a positive Bernoulli constant, which means that the matter in the post-shock region is able to escape to infinity. However, positiveness of the Bernoulli constant may lead to another situation as well, where the shock may quasi-periodically originate at some certain radius and propagate outwards without the formation of outflows. So the formation of outflows is one of the possible scenarios when we focus on the positive energy solutions. In this paper we concentrate only on solutions producing outflows, and we thus use only positive Bernoulli constant throughout our model, which is standard practice for studying hydrodynamic winds. The presence of such a collisionless, steady, standing, spherical shock as discussed above may randomize the directed infall motion, and at the shock surface the individual components of the total energy of the flow get rearranged in such a manner that the thermal energy term dominates (due to shock-generated enormous post-shock proton temperature) over the gravitational attraction of the accretor, and a part of the infalling material is driven by thermal pressure to escape to infinity as wind. It is assumed that the outflow has exactly the same temperature as the post-shock flow temperature of protons, but the energy is not conserved, as matter jumps from the accretion to the wind branch. In other words, the outflow is assumed to be kept in a thermal bath of temperature as that of the post-shock accretion flows. At the shock, entropy is generated, and the outflow as well as the post-shock inflow will have higher entropy density for the same specific energy. It is also assumed that the polytropic index $\gamma$ of the *inflow* can be varied but that of the outflow is always unity. Another assumption made is to treat the accreting as well as the post-shock matter as a single-temperature fluid, the temperature of which is basically characterized by proton temperature.

The plan of this paper is as follows. In the next section, we describe our model and present the set of equations governing the polytropic inflow and isothermal outflow along with their simultaneous solution scheme. In Section 3, we present our results. Finally, in Section 4, the results will be reviewed and conclusions will be drawn.

## 2 MODEL DESCRIPTION, GOVERNING EQUATIONS AND SOLUTION PROCEDURE

### 2.1 Inflow model

We assume that a Schwarzschild-type black hole quasi-spherically accretes fluid obeying the polytropic equation of state. The density of the fluid is $\rho(r)$, $r$ being the radial distance measured in units of the Schwarzschild radius $r_{\rm g}$. We also assume that the accretion rate (in units of the Eddington rate $\dot{M}_{\rm Edd}$) is not a function of $r$. We ignore the self-gravity of the flow, and carry out the calculation using the Paczyński–Wiita (Paczyński & Wiita 1980) potential, which mimics the surroundings of a Schwarzschild black hole. Also effects due to magnetic field have been ignored. The equations (in dimensionless geometric units) governing the inflow are (for details see D99a):

(a) conservation of specific energy

$$\mathcal{E} = \frac{u(r)^2}{2} + na(r)^2 - \frac{1}{2(r-1)}, \quad (1)$$

(b) mass conservation

$$\dot{M}_{\text{in}} = \Theta_{\text{in}} \rho(r) u(r) r^2, \quad (2)$$

where $\Theta_{\text{in}}$ is the solid angle subtended by the inflow.

In this work we normally use the value of polytropic index $\gamma$ of the inflow to be 4/3. Though far away from the black hole, optically thin accreting plasma may not be treated as a relativistic fluid in general; close to the hole it *always* advects with enormously large radial velocity and could be well approximated as relativistic with $\gamma = 4/3$. As our main region of interest, the shock formation zone, *always* lies close to the black hole (a few tens of $r_g$ away from the hole or sometimes even less; see results and figures in Section 3), we believe that it is fairly justifiable to assign the value 4/3 for $\gamma$ in our work. However, to model a real flow rigorously without any assumptions, a variable polytropic index having the proper functional dependence on radial distance [i.e. $\gamma \equiv \gamma(r)$] might be considered instead of using a constant $\gamma$, and the equations of motion might be formulated accordingly, which we did not attempt here for the sake of simplicity. Nevertheless, we keep our options open for values of $\gamma$ other than 4/3 as well, and investigate the outflow for a range of values of $\gamma$ of the *inflow* for a specific value of $\mathcal{E}$ and $\dot{M}_{\text{Edd}}$ (see Fig. 4, Section 3.2.3). The same kind of investigations could be performed for a variety of values of $\mathcal{E}$ and $\dot{M}_{\text{Edd}}$ and a set of results may be obtained with a wide range of values of $\gamma$, which shows that our calculation is not restricted to the value $\gamma = 4/3$. Rather, the model is general enough to deal with all possible values of $\gamma$ for polytropic accretion.

As already mentioned, a steady, collisionless shock forms as a result of the instabilities in the plasma flow (PK83; KE86). We assume that, for our model, the effective thickness of the shock $\Delta_{\text{sh}}$ is small enough compared to the shock stand-off distance $r_{\text{sh}}$, i.e. that

$$\Delta_{\text{sh}} \ll r_{\text{sh}},$$

and that the relativistic particles encounter full shock compression ratio while crossing the shock. At the shock, the density of matter will shoot up and the inflow velocity will fall abruptly. If $(\rho_-, u_-)$ and $(\rho_+, u_+)$ are the pre- and post-shock densities and velocities respectively at the shock, then

$$\frac{\rho_+}{\rho_-} = R_{\text{comp}} = \frac{u_-}{u_+}, \quad (3)$$

where $R_{\text{comp}}$ is the shock compression ratio. For high shock Mach number solution (which is compatible with our low-energy accretion model), the expression for $R_{\text{comp}}$ can be well approximated as

$$R_{\text{comp}} = 1.44 M_{\text{sh}}^{3/4}, \quad (4)$$

where $M_{\text{sh}}$ is the shock Mach number; equation (4) holds for $M_{\text{sh}} \gtrsim 4.0$ (Ellison & Eichler 1985).

Shock location may be computed as (see D99a for details):

$$r_{\text{sh}} = \frac{3\sigma_{\text{pp}} c \dot{M}_{\text{in}}}{u_{\text{sh}}^2 \Theta_{\text{in}}} \left( \frac{\delta}{\mathcal{E}_F} \right). \quad (5)$$

The ratio $(\delta/\mathcal{E}_F)$ as a function of shock Mach number $M_{\text{sh}}$ for high shock Mach number solution (low-energy inflow) is obtained from the empirical solution deduced by Ellison & Eichler (1984), after suitable modification required for our model.

### 2.2 Outflow model

As the isothermality assumption for the outflow (at least up to the sonic point) has been justified as a valid one (see 'Introduction'), the equation of state for the outflowing matter can be written as

$$P = \frac{R}{\mu} \rho T = C_s^2 \rho, \quad (6)$$

where $R$ is the universal gas constant, $\mu$ is the mean molecular weight, $T$ is the temperature, $\rho$ is the density and $C_s$ is the isothermal sound speed of the outflow. Because of the fact that the subsonic outflow is taken to originate from the shock surface, $T$ in equation (6) is basically the post-shock temperature.

Using equation (6) we integrate the radial momentum conservation equation

$$v(r) \frac{dv(r)}{dr} + \frac{1}{\rho(r)} \frac{dP(r)}{dr} + \frac{1}{(r-1)^2} = 0 \quad (7)$$

and continuity equation

$$\frac{1}{r^2} \frac{d}{dr} [\Theta_{\text{out}} \rho(r) v(r)^2] = 0 \quad (8)$$

to obtain the following two conservation equations for our outflow:

$$\frac{v_{\text{iso}}^2}{2} + C_s^2 \ln \rho_{\text{iso}}(r) - \frac{1}{2(r-1)^2} = \text{constant} \quad (9)$$

and

$$\dot{M}_{\text{out}} = \Theta_{\text{out}} \rho_{\text{iso}}(r) v_{\text{iso}}(r) r^2. \quad (10)$$

The subscript 'iso' indicates that all the relevant physical parameters are measured based on the isothermality assumption. $\dot{M}_{\text{out}}$ is the quantity of outflowing mass, which is constant for a fixed value of our three-parameter input set $\{P_3\}$, whereas $v_{\text{iso}}(r)$ is the radial velocity of the outflow and $\Theta_{\text{out}}$ is the solid angle subtended by the outflow.

The isothermal sound speed $C_s$ (measured at the shock surface) and the post-shock temperature $T_{\text{sh}}$ can be written as

$$C_s = \sqrt{\frac{p_+(r_{\text{sh}})}{\rho_+(r_{\text{sh}})}} \quad (11)$$

and

$$T_{\text{sh}} = \frac{C_s^2 \mu m}{\kappa}, \quad (12)$$

where $p_+(r_{\text{sh}})$ and $\rho_+(r_{\text{sh}})$ are the post-shock pressure and density at the shock surface, $\mu$ is the mean molecular weight, $m$ is the mass of each particle and $\kappa$ is Boltzmann's constant. The temperature of the outflow in our work will essentially be assumed to be characterized by proton temperature. Hence we replace $m$ by $m_p$ (mass of the proton) and take $\mu = 0.5$ for our calculation. Hence the post-shock proton temperature would be

$$T_{\text{psh}} = \frac{C_s^2 \mu m_p}{\kappa}. \quad (13)$$

As the total pressure (thermal pressure plus ram pressure) of the polytropic inflow is a shock-invariant quantity, we can write

$$p_+(r_{sh}) + \rho_+(r_{sh})u_+^2(r_{sh}) = p_-(r_{sh}) + \rho_-(r_{sh})u_-^2(r_{sh}), \quad (14)$$

where the $+$ and $-$ denote the post- and pre-shock quantities respectively. For low-energy cold inflow, the thermal pressure of the pre-shock accreting material is negligible compared to its ram pressure. Hence using equation (3), equation (14) can be approximated as

$$p_+(r_{sh}) = u_{sh}^2 r_{sh}\left(\frac{R_{comp} - 1}{R_{comp}}\right). \quad (15)$$

Using the above value of $p_+(r_{sh})$ and equations (3) and (4), it is easy to express isothermal sound speed $C_s$ and post-shock proton temperature (which is being treated as the effective characteristic outflow temperature) $T_{psh}$ (equations 11 and 13) in terms of accretion rate and various shock parameters as

$$C_s = 0.694 M_{sh}^{-0.75}\sqrt{\frac{u_{sh}^3 r_{sh}^3}{\dot{M}_{in}}(1.44 M_{sh}^{0.75} - 1)} \quad (16)$$

and

$$T_{psh} = \frac{0.24 M_{sh}^{-1.5} m_p u_{sh}^3 r_{sh}^3}{\kappa \dot{M}_{in}}(1.44 M_{sh}^{0.75} - 1). \quad (17)$$

As the outflow is taken to be isothermal, $T_{psh}$ will be constant throughout the flow (at least up to the sonic point, up to which isothermality is a sufficiently valid assumption). Also it is to be noted that from equation (13), as $\kappa$ and $m_p$ are constant, the outflow sound speed will also be constant for a fixed value of $\{P_3\}$. However, as all the shock parameters $(u_{sh}, r_{sh}, M_{sh})$ are derivable from $\{P_3\}$, both $C_s$ and $T_{psh}$ will vary with the variation of $\{P_3\}$ for polytropic inflow.

For simplicity of calculation, we assume that the outflow is also quasi-spherical and $\Theta_{out} \approx \Theta_{in}$. Defining $R_{\dot{m}}$ as the mass outflow rate, we obtain

$$R_{\dot{m}} = \dot{M}_{out}/\dot{M}_{in}. \quad (18)$$

### 2.3 Procedure to solve the inflow and outflow equations simultaneously

Having thus computed the outflow temperature and sound speed at the shock surface, our next step is to calculate the value of $\rho_{iso}(r)$ and $v_{iso}(r)$ and to use these values to get the value of $\dot{M}_{out}$ in geometric units and to obtain the value of $R_{\dot{m}}$ from equation (18) using the values of $\dot{M}_{in}$ and $\dot{M}_{out}$.

However, in this paper, as we are interested to find out the *ratio* of $\dot{M}_{out}$ to $\dot{M}_{in}$ (equation 18), and *not* the explicit value of $\dot{M}_{out}$, individual calculation of $\Theta_{out}$ and $\Theta_{in}$ is *not* required, since the fraction $\Theta_{out}/\Theta_{in}$ (appearing in the expression for $R_{\dot{m}}$ via $\dot{M}_{out}$ and $\dot{M}_{in}$) becomes unity owing to our simplified assumption of flow geometry (i.e. $\Theta_{out} \approx \Theta_{in}$).

Before we proceed into further detail, a general understanding of the transonic inflow–outflow system in the present case is essential to understand the basic scheme of the solution procedure. Let us consider the transonic accretion first. Infalling matter becomes supersonic after crossing a saddle-type sonic point, the location of which is determined by $\{P_3\}$. This supersonic flow then encounters a shock (if present), the location of which $(r_{sh})$ is determined from equation (5). At the shock surface, part of the incoming matter, having higher entropy density (because a shock in a fluid flow generates entropy), is likely to return back as wind through a sonic point *other than* the one through which it just entered. Thus a combination of transonic topologies – one for the polytropic inflow and another for the isothermal outflow (passing through a *different* sonic point and following a topology *completely different* from that of the 'self-wind' of the accretion) – is required to obtain a full solution. So it turns out that finding a complete set of self-consistent inflow–outflow solutions ultimately boils down to locating the sonic point of the *isothermal outflow* and the mass flux through it. Thus a supply of parameters $\mathcal{E}$, $\dot{M}_{in}$ (in units of $\dot{M}_{Edd}$) and $\gamma$ makes a self-consistent computation of $R_{\dot{m}}$ possible.

The following procedure is adopted to obtain a complete simultaneous solution of inflow–outflow equations: From equations (1) and (2), the rate of change of polytropic radial velocity $u(r)$ of the *inflow* is obtained as

$$\frac{du(r)}{dr} = \frac{2a(r)^2/r - 1/[2(r-1)^2]}{u(r) - a(r)^2/u(r)}, \quad (19)$$

where, as already mentioned, $a(r)$ is the polytropic sound speed, which is a function of the radial distance measured from the black hole in units of the Schwarzschild radius. At the sonic point, the numerator and denominator separately vanish and give rise to the so-called sonic point condition:

$$u_c = a_c = \frac{\sqrt{r_c}}{2(r_c - 1)}, \quad (20)$$

where the subscript $c$ represents the quantities at the sonic point. The derivatives at the sonic point $(du/dr)_c$ are computed by applying L'Hospital's rule on equation (19). The expression for $(du/dr)_c$ is obtained by solving the following polynomial:

$$\left(\frac{2n+1}{n}\right)\left(\frac{du}{dr}\right)_c^2 + \frac{3u_c}{nr_c}\left(\frac{du}{dr}\right)_c$$
$$- \left[\frac{1}{(r_c - 1)^3} - \frac{2a_c^2}{r_c}\left(\frac{n+1}{n}\right)\right] = 0. \quad (21)$$

We obtain the inflow sonic point $r_c$ by solving equations (1) and (20) as

$$r_c = \frac{2\mathcal{E} - 1}{2\mathcal{E} - \left(n + \frac{1}{2}\right)}.$$

Using the fourth-order Runge–Kutta method, $u(r)$, $a(r)$ and the inflow Mach number $u(r)/a(r)$ are computed along the inflow from the *inflow* sonic point $r_c$ till the position where the shock forms. The shock location is calculated by simultaneously solving equations (1), (2) and (5). Various shock parameters (i.e. density, pressure, etc., at the shock surface) are then computed self-consistently. The shock compression ratio $R_{comp}$, isothermal sound speed $C_s$ and the effective characteristic outflow temperature $T_{psh}$ are then computed using equations (4), (16) and (17) respectively. Using equations (9) and (10), we now express the derivative of the *outflow* velocity with respect to the radial distance as

$$\frac{dv_{iso}(r)}{dr} = \frac{2C_s^2/r - 1/[2(r-1)^2]}{v_{iso}(r) - C_s^2/v_{iso}(r)}. \quad (22)$$

As has been done for the case of polytropic inflow, here also we obtain the sonic point conditions for the outflow from the requirement that the numerator and denominator of equation (22) would vanish separately. The two sonic point conditions are thus

expressed as

$$r_{iso}^2|_c - r_{iso}|_c \left(2 + \frac{1}{4C_s^2}\right) + 1 = 0 \tag{23}$$

and

$$v_{iso}|_c = C_s. \tag{24}$$

Equation (23) also gives the location of the outflow sonic point $r_{iso}|_c$ as the value of $C_s$ is known (by solving equation 16) for a fixed set of $\{P_3\}$. The derivative of the flow velocity at $r_{iso}|_c$ [we call it the 'critical derivative' and denote it by $(dv_{iso}/dr)_c$] is computed using L'Hospital's rule on equation (22) and can be obtained by solving the following equation:

$$\left(\frac{dv_{iso}}{dr}\right)_c^2 + \left(\frac{dv_{iso}}{dr}\right)_c + \left[\frac{2C_s^2}{r_c^2} - \frac{1}{(r_c-1)^3}\right] = 0. \tag{25}$$

The fourth-order Runge–Kutta method is then employed to integrate equation (22) to find the outflow velocity and density at the shock location. The outflow rate is then computed using equation (18).

## 3 RESULTS

### 3.1 Flow topology

Fig. 1 shows a typical solution which combines the accretion and the outflow. Mach number is plotted along the Y axis, and the distance (in geometric units) from the event horizon of the accreting black hole is plotted along the X axis in logarithmic scale. The input parameters are $\mathcal{E} = 0.0001$, $\dot{M}_{in} = 0.2$ Eddington rate ($\dot{M}_{Edd}$ stands for the Eddington rate in the figure) and $\gamma = 4/3$ corresponding to relativistic inflow. The solid curve AB represents the pre-shock region of the inflow, and the solid vertical line BC with double arrow at $r_{sh}$ represents the shock transition. The location of the shock ($15.92r_g$) is obtained using equation (5) for the particular set of inflow parameters mentioned above. The dotted curve CC'D represents the isothermal outflow branch which is generated, according to our model, from the shock surface. The outflow sonic point $S_{out}$ (marked by ⊙ at C' on CC'D) comes out to be $35.21r_g$. After crossing the sonic point, outflow becomes supersonic and this supersonic wind C'D extends away to the vastness of intergalactic space. Notice that the outflow sonic point $S_{out}$ ($35.21r_g$) and solution topology CC'D are *completely different* from that of the 'self-wind' of accretion flow $S_{in}$ ($3753.33r_g$) and EF. This is due to the fact that outflow comes with shock-generated higher entropy density and the outflowing matter is assumed to obey the isothermal equation of state. It is also observed that the outflow sonic point is, in general, located *closer* to the event horizon compared to the inflow sonic point. The mass outflow rate $R_{\dot{m}}$ in the case shown in Fig. 1 is 0.05.

### 3.2 Dependence of $R_{\dot{m}}$ on $\{P_3\}$

#### 3.2.1 Variation of $R_{\dot{m}}$ with $\mathcal{E}$

In Fig. 2(a), we have plotted the variation of $R_{\dot{m}}$ with inflow specific energy $\mathcal{E}$ for a set of equispaced values of $\dot{M}_{in}$ (measured in units of Eddington rate $\dot{M}_{Edd}$ as marked in the figure). The unequal gaps between the curves imply that, when $\mathcal{E}$ is kept fixed, $R_{\dot{m}}$ *anticorrelates non-linearly* with $\dot{M}_{in}$ (which is explicitly manifested in Fig. 3a). It is interesting to note that different curves terminate at different points, which indicates that the maximum energy $\mathcal{E}_{max}$ of the *inflow* for which the shock forms, increases non-linearly with the decrease of Eddington rate. This implies that a pair-plasma pressure-mediated shock does *not* form for all values of $\{P_3\}$; rather, a specific region of parameter space spanned by $\mathcal{E}$, $\dot{M}_{in}$ and $\gamma$ allows shock formation. This fact will be further supported from the next set of figures where the dependence of $R_{\dot{m}}$ with $\dot{M}_{in}$ is shown explicitly.

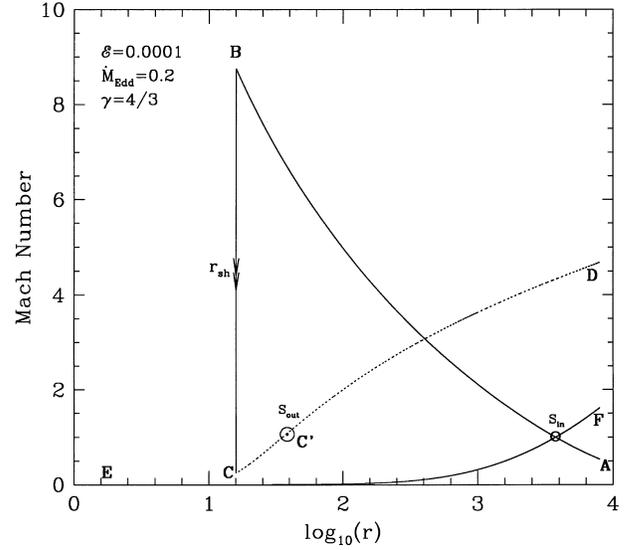

**Figure 1.** Combined solution topology for hybrid transonic inflow–outflow system. Solid line AB represents polytropic accretion (up to the shock location, which is at $15.92r_g$), while the dotted line CC'D is for isothermal outflow. Solid vertical line BC with the double arrow $r_{sh}$ represents the shock transition. $S_{out}$ ($35.21r_g$) and $S_{in}$ ($3753.33r_g$) are the sonic points for the outflow and the inflow, respectively, and the solid curve EF represents the 'self-wind' of the polytropic accretion. Inflow parameters used for this case are $\mathcal{E} = 0.0001$, $\dot{M}_{in} = 0.2\dot{M}_{Edd}$ and $\gamma = 4/3$. Mass outflow rate $R_{\dot{m}}$ here comes out to be 0.05. (See text for details.)

From Fig. 2(a) it is clear that, for a fixed value of $\dot{M}_{in}$, $R_{\dot{m}}$ increases with $\mathcal{E}$ monotonically and non-linearly. This is because, as $\mathcal{E}$ increases keeping the accretion rate constant, the shock Mach number $M_{sh}$ decreases, the result of which is the decrement of shock location $r_{sh}$ but increment of post-shock density $\rho_+(r_{sh})$, pressure $p_+(r_{sh})$ and temperature $T_{psh}$. The higher the post-shock proton temperature, the higher is the isothermal sound speed $C_s$ of the outflow, as well as the isothermal bulk velocity $v_{iso}(r_{sh})$ with which the outflow leaves the shock surface. The outflow rate, which is the product of three quantities, namely $r_{sh}$, $\rho_+(r_{sh})$ and $v_{iso}(r_{sh})$ (see equation 10), increases in general due to the combined 'tug-of-war' of these three quantities. Moreover, the closer the shock is to the black hole, the greater will be the amount of gravitational potential available to be put on to the relativistic protons to provide stronger outward pressure; and the closer the shock forms to the black hole, the higher will be the post-shock proton temperature (the effective characteristic outflow temperature) and the higher will be the amount of outflow (as isothermal wind is highly thermally driven, a fact that will be more strongly established in the next set of results). The dependence of mass outflow rate in this case (variable energy with constant accretion rate) on the various parameters mentioned above is manifested in Fig. 2(b), where we have shown the variation of $R_{\dot{m}}$ with different shock parameters for a fixed accretion rate ($1.0\dot{M}_{Edd}$) but with variable energy. The solid curve marked with $\rho_+(r_{sh})$ represents the variation of post-shock density; the dotted curve marked with $r_{sh}$, the short-dashed curve marked with $T_{psh}$ and long-dashed curve marked with $V_{iso}(r_{sh})$ represent the variation of shock location, post-shock proton temperature and the launching velocity of outflow at the shock surface, respectively. The $\gamma$ value of the inflow is taken to be 4/3. While plotting, different shock parameters are scaled as follows: $r_{sh} \to 0.5 r_{sh}$, $\rho_+(r_{sh}) \to \rho_+(r_{sh}) \times 10^{22}$, $T_{psh} \to T_{psh} \times 10^{10}$ and $V_{iso}(r_{sh}) \to V_{iso}(r_{sh}) \times 363.64$.

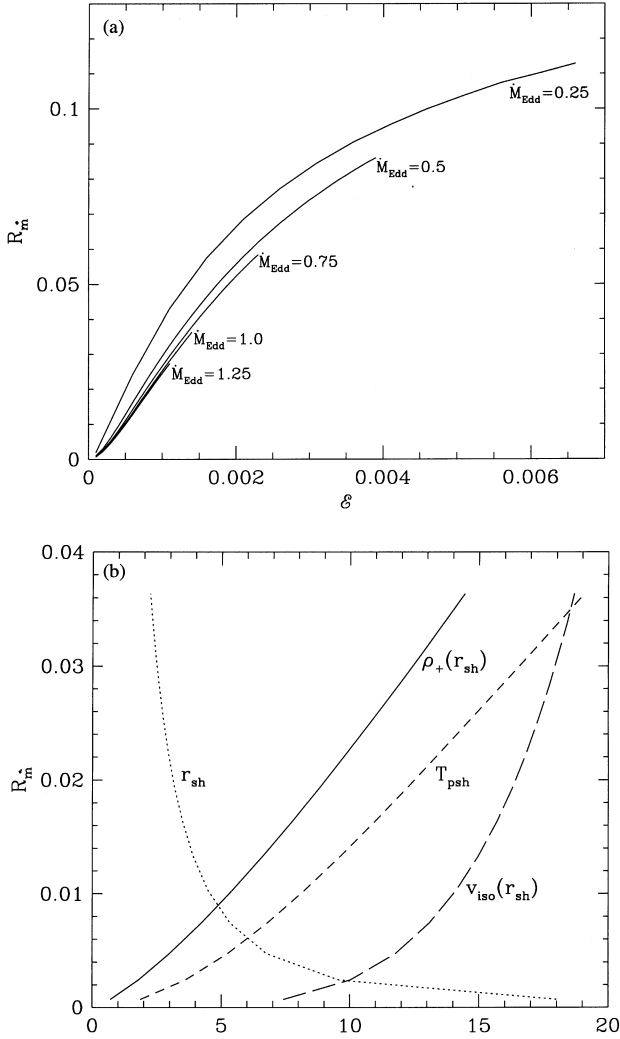

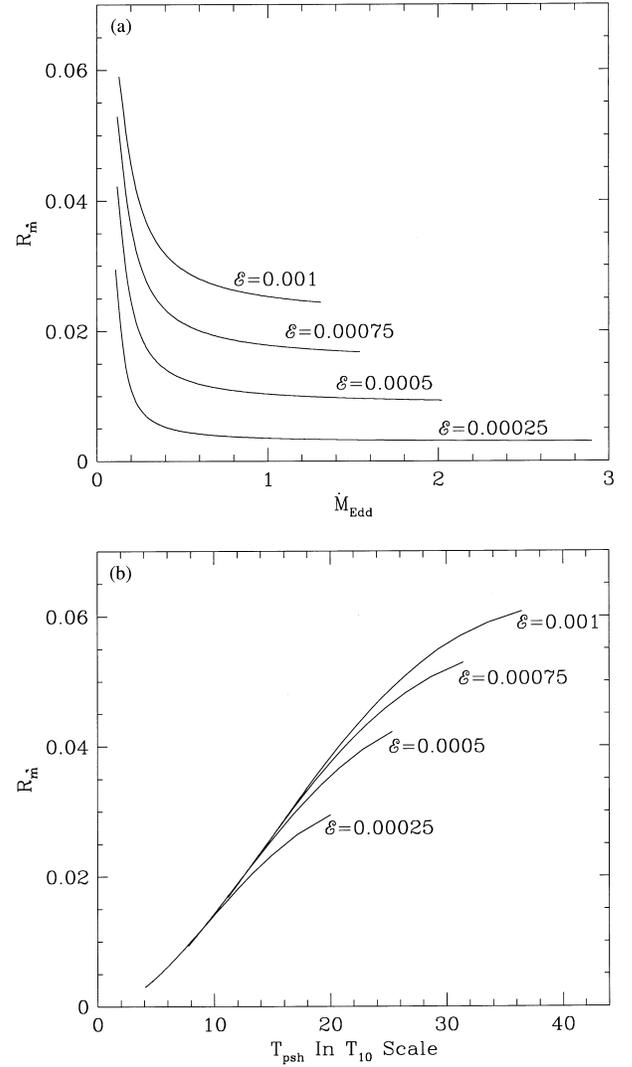

**Figure 2.** (a) Variation of $R_{\dot m}$ as a function of the specific energy $\mathcal{E}$ of the polytropic accretion for different Eddington rates (marked as $\dot M_{Edd}$ in the figure). The maximum energy $\mathcal{E}_{max}$ for which the shock forms, increases non-linearly with decrease of accretion rate. $\gamma$ of the inflow is taken to be 4/3. (b) Variation of $R_{\dot m}$ with different shock parameters for a fixed accretion rate ($1.0\dot M_{Edd}$) but with variable energy. Solid curve marked with $\rho_+(r_{sh})$ represents the variation of post-shock density; the dotted curve marked with $r_{sh}$, the short-dashed curve marked with $T_{psh}$ and long-dashed curve marked with $V_{iso}(r_{sh})$ represent the variation of shock location, post-shock proton temperature and the launching velocity of outflow at the shock surface, respectively. $\gamma$ of the inflow is taken to be 4/3. (See text for details.) In the plotting, different shock parameters are scaled as follows: $r_{sh} \to 0.5 r_{sh}$, $\rho_+(r_{sh}) \to \rho_+(r_{sh}) \times 10^{22}$, $T_{psh} \to T_{psh} \times 10^{10}$ and $V_{iso}(r_{sh}) \to V_{iso}(r_{sh}) \times 363.64$.

As a low-energy solution corresponds to a high shock Mach number, and as $R_{\dot m}$ increases with $\mathcal{E}$ (for constant $\dot M_{in}$), an inflow with *low* shock Mach number [and hence with low shock compression ratio $R_{comp}$ via equation (4)] produces more outflow.

**Figure 3.** (a) Variation of $R_{\dot m}$ as a function of accretion rate (in units of Eddington rate $\dot M_{Edd}$) for a set of values of specific energy $\mathcal{E}$ of the inflow (marked on the figure). It is observed that $R_{\dot m}$ anticorrelates non-linearly with accretion rate, which tells us that low-luminosity objects produce more outflows. The value of $\gamma$ is taken to be 4/3. (b) Variation of $R_{\dot m}$ with post-shock proton temperature $T_{psh}$ (scaled in units of $T_{psh}^{10}$ and marked as $T_{10}$ in the figure) corresponding to the data set of part (a). (See text for details.)

### 3.2.2 Variation of $R_{\dot m}$ with $\dot M_{in}$

In Fig. 3(a), we show the outflow rate $R_{\dot m}$ as a function of the accretion rate (measured in terms of Eddington rate $\dot M_{Edd}$) of the incoming flow for a range of fixed specific energy $\mathcal{E}$ of the inflow. $\mathcal{E}$ corresponding to each curve is marked in the figure. It is interesting to note that the outflow occurs for both super- and sub-Eddington accretion and low-luminosity objects produce more outflow as is observed. It is also to be noted in this context that the maximum value of accretion rate $\dot M_{Edd}|_{max}$ for which the shock forms, increases with decrease of the inflow specific energy $\mathcal{E}$. This figure is drawn for $\gamma = 4/3$.

In Fig. 3(b), we plot the variation of $R_{\dot m}$ directly with the post-shock proton temperature $T_{psh}$ corresponding to the accretion rate shown in Fig. 3(a). Post-shock proton temperature $T_{psh}$ (in units of $T_{psh}^{10}$, which we mark as $T_{10}$ in the figure) is plotted along the X axis. As is observed from the figure, the outflow is clearly thermally driven. Hotter flow produces more winds as is expected. From Figs 3(a) and (b), it is clear that a 'high energy–low luminosity' combination for polytropic accretion maximizes the post-shock proton temperature and thus gives rise to the highest amount of outflows.

It is observed that, while the shock location decreases with decrease in accretion rate (for a fixed energy), the launching velocity of outflow at the shock surface $v_{iso}(r_{sh})$, the post-shock density $\rho_+(r_{sh})$ and the temperature of the flow increase. It is also observed that the shock Mach number $M_{sh}$ as well as the shock compression ratio $R_{comp}$ increase with decrease in accretion rate. So for varying accretion rate for a fixed energy, high shock Mach number solutions produce higher outflow and the value of $R_{comp}$ is also necessarily high to obtain high $R_{\dot{m}}$. This is in contrast to the results obtained in Section 3.2.1 (dependence of $R_{\dot{m}}$ on $\mathcal{E}$ keeping $\dot{M}_{in}$ constant) where low shock Mach number solutions produced higher outflow rates and the value of $R_{comp}$ was also low for that case. So in some sense, while varying accretion rate for a fixed $\mathcal{E}$ gives high outflow rates for 'strong shock' cases (i.e. high shock Mach number with strong shock compression of matter), variation of inflow specific energy $\mathcal{E}$ (keeping accretion rate $\dot{M}_{in}$ fixed) prefers 'weak shock' solutions (low $M_{sh}$ and low shock compression of matter) to produce high outflow.

### 3.2.3 Variation of $R_{\dot{m}}$ with $\gamma$

In previous cases, the polytropic index $\gamma$ of the accreting matter was always kept fixed at the value 4/3. To have a better insight into the behaviour of the outflow, we plot $R_{\dot{m}}$ as a function of $\gamma$ (Fig. 4) for a fixed value of $\mathcal{E} = 0.0001$ and $\dot{M}_{in} = 0.2\dot{M}_{Edd}$. The range of $\gamma$ shown here is the range for which a shock forms for the specified values of $\mathcal{E}$ and $\dot{M}_{in}$. It is observed that $R_{\dot{m}}$ correlates with $\gamma$. We have found that the outflow temperature increases with increase in $\gamma$ while $M_{sh}$ and $R_{comp}$ decrease. So here also the flow is essentially thermally driven and 'weak shock' solutions are preferred to obtain high outflows.

### 3.3 Temperature dependence of $R_{\dot{m}}$: a generic result

In *all* of the previous examples, where we have studied the dependences of $R_{\dot{m}}$ on various input parameters as well as shock parameters, it is observed that whether the mass outflow rate correlates or anticorrelates with any of the members of {$P_3$} (our three-parameter input set) or with any shock parameters ($r_{sh}$, $M_{sh}$ or $R_{comp}$, for example), $R_{\dot{m}}$ *always correlates* with the outflow temperature, which tells us that isothermal outflow is essentially thermally driven *in general*. That means that if a shock forms, whatever the initial flow conditions and whatever the nature of the dependence of $R_{\dot{m}}$ on any of the inflow/shock parameters, hotter flows *always* produce more winds. Hence the post-shock temperature plays a crucial role in our model. The above fact is manifested in Figs 5(a) and (b). In Fig. 5(a), we plot $R_{\dot{m}}$ directly with post-shock proton temperature obtained for three different cases. The solid line marked with $\mathcal{E}$ gives the temperature changes for varying inflow specific energy while keeping accretion rate and polytropic index constant; the dotted line marked with $\dot{M}_{Edd}$ is drawn for the case where the accretion rate (in units of Eddington rate) is varied for constant values of energy and polytropic index; and lastly, the dashed line marked with $\gamma$ is obtained for variation of the polytropic index of the inflow keeping energy and accretion rate constant. It is clear from the figure that $R_{\dot{m}}$ *always* increases with increase of post-shock proton temperature (thus with the outflow temperature). In Fig. 5(b), we directly show the dependence of post-shock proton temperature on $\mathcal{E}$, $\dot{M}_{in}$ and $\gamma$ of the inflow. The solid curve gives the variation with energy, while the dotted and the dashed curves represent the variation with accretion rate and polytropic index of the inflow, respectively. It is observed that, while flow temperature increases with increase of $\mathcal{E}$ and $\gamma$, decrease of $\dot{M}_{in}$ (measured in units of Eddington rate and marked as $\dot{M}_{Edd}$ in the figure) increases the flow temperature $T_{psh}$. The inference drawn by combining Figs 5(a) and (b) tells us that, while $R_{\dot{m}}$ correlates with $\mathcal{E}$ and $\gamma$, it anticorrelates with the accretion rate, which again is in one-to-one agreement with the results obtained in Sections 3.2.1–3.2.3.

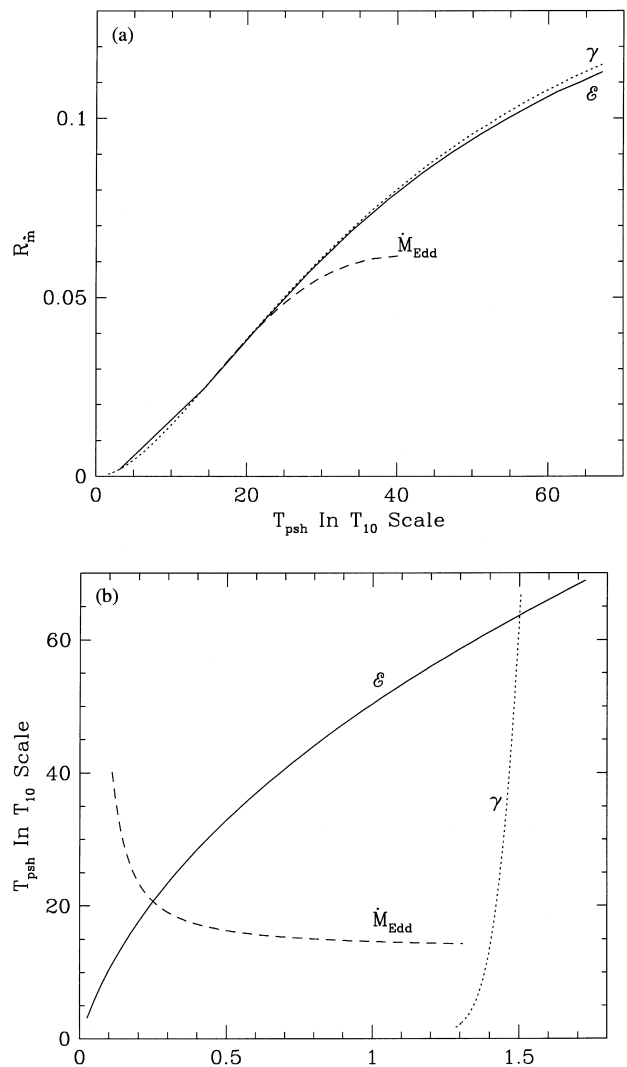

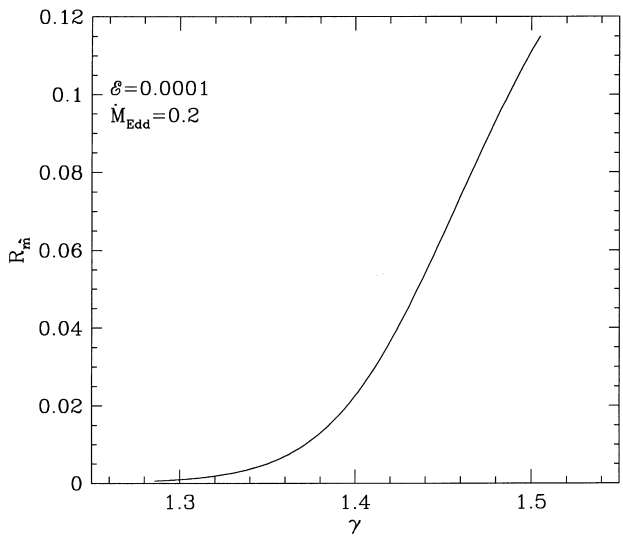

**Figure 4.** Variation of $R_{\dot{m}}$ with polytropic index $\gamma$ of the inflow for a fixed value of specific energy $\mathcal{E} = 0.0001$ and accretion rate $\dot{M}_{in} = 0.2\dot{M}_{Edd}$. It is observed that $R_{\dot{m}}$ in general correlates with $\gamma$. (See text for details.)

**Figure 5.** (a) Direct variation of $R_{\dot{m}}$ with post-shock proton temperature $T_{psh}$ (scaled in units of $T_{psh}^{10}$ and marked as $T_{10}$ in the figure) for variation of specific energy $\mathcal{E}$ (solid curve), accretion rate in terms of Eddington rate (dashed curve) and polytropic index $\gamma$ (dotted curve) of the inflow. (b) Dependence of post-shock proton temperature $T_{psh}$ (scaled in units of $T_{psh}^{10}$ and marked as $T_{10}$ in the figure) with specific energy $\mathcal{E}$ (solid curve), accretion rate in terms of Eddington rate (dashed curve) and polytropic index $\gamma$ (dotted curve) of the inflow. In the plotting $\mathcal{E}$ is scaled as $\mathcal{E} \rightarrow 250.0 \times \mathcal{E}$.

Our calculations in this paper, being simply founded, do not explicitly include various radiation losses and cooling processes, the combined effects of which may reduce the post-shock proton temperature and the outflow temperature. So, in reality, these could be lower than what we have obtained here, and the amount of outflow would be less than what is obtained in our calculation. This deviation will be more important for the cases with high accretion rate. Nevertheless, cases for low accretion rate discussed here would not be affected that much, and our preliminary investigation shows that, even if we incorporate various losses, the overall profile of the various curves showing the dependence of $R_{\dot{m}}$ on different inflow parameters would be *exactly* the same, only the numerical value of $R_{\dot{m}}$ in some cases (especially for high accretion) might decrease. Detailed computation of $R_{\dot{m}}$ including various losses is in progress and will be presented elsewhere.

## 4 CONCLUDING REMARKS

Although there have been many attempts in the literature to study the astrophysical outflows from Galactic and extragalactic sources, till now, to the best of our knowledge, no such model is available that can compute the mass outflow rate from non-rotating accretion of this kind in terms of *only three* inflow parameters and is able to study rigorously the dependence of $R_{\dot{m}}$ on various physical parameters governing the accretion. We believe that this has been successfully performed in our present work. We take a spherical, steady, self-supported pair-plasma pressure-mediated shock surface around a Schwarzschild black hole as the 'effective' physical barrier from where the outflow could be generated. We then calculate the location of this shock surface and compute the rate of the outflow generated from this surface in terms of *only three inflow parameters*, namely, the specific energy $\mathcal{E}$, the mass accretion rate $\dot{M}_{in}$ and the polytropic index $\gamma$, and rigorously study the dependence of this rate on various physical parameters governing the inflow and different shock parameters. We take polytropic accretion, while the outflow is taken to be isothermal. We thus self-consistently construct a 'hybrid' inflow–outflow model which could successfully connect flows obeying two different equations of state.

At this point, it is worth mentioning that the hot and dense shock surface around black holes, which served here as the effective physical barrier around compact objects regarding the mass outflow, may be generated due to other physical effects as well for spherical accretion (Meszaros & Ostriker 1983; Chang & Ostriker 1985; Babul, Ostriker & Meszaros 1989; Park 1990a,b). Another very important approach launched recently to construct such an 'effective barrier' for non-spherical flows (for rotating flows forming the accretion discs) was to introduce the concept of centrifugal pressure-supported boundary layers (CENBOL). Treating the CENBOL as the effective atmosphere of the rotating flows around compact objects [which forms as a result of a standing Rankine–Hugoniot shock or due to the maximization of polytropic pressure of accreting material in the absence of a shock (see Das 1998; Das & Chakrabarti 1999 and references therein)], detailed computation of mass outflow rate from the advective accretion discs has been done, and the dependence (both explicit and implicit) of this rate on various accretion and shock parameters has been studied quantitatively very recently (Das 1998, 1999c; Das & Chakrabarti 1999) by constructing a self-consistent disc–outflow system. Also the possibility of outflow-driven contamination of metallicity to the outer galaxies has been discussed there. A similar kind of problem (outflows from CENBOL) has also been treated in a rather qualitative way to indicate the effects of outflow on the spectral states and Quasi Periodic Oscillation (QPO) properties of black hole candidates (Chakrabarti 1999).

The basic conclusions of this paper are the following:

(i) It is possible to construct a 'hybrid' inflow–outflow system (which could connect flows with two different equations of state) to compute the mass outflow rate $R_{\dot{m}}$ from matter with negligible or almost zero intrinsic angular momentum accreting on a Schwarzschild black hole.

(ii) The pair-plasma pressure-mediated shock surface can serve as the 'effective' physical barrier around the black hole regarding the computation of mass outflow rate.

(iii) Computation of $R_{\dot{m}}$ can be carried out in terms of only three inflow parameters: $\{\mathcal{E}, \dot{M}_{in}, \gamma\}$.

(iv) Mass outflow rate can vary anywhere from a fraction of one per cent to a couple of per cent depending on parameters of $\{P_3\}$; thus a wide range of $R_{\dot{m}}$ has been investigated successfully.

(v) A shock does not always form. Although for most regions of parameter space (spanned by $\mathcal{E}$, $\dot{M}_{in}$ and $\gamma$) a shock does form, a shock does not form for all values of $\mathcal{E}$, $\dot{M}_{in}$ and $\gamma$.

(vi) While $R_{\dot{m}}$ increases with increase in $\mathcal{E}$ and $\gamma$, a decrease in $\dot{M}_{in}$ increases $R_{\dot{m}}$. In other words, low-luminosity objects produce high outflow and the 'high energy–low accretion rate' solutions are the best choice to maximize $R_{\dot{m}}$.

(vii) Isothermal outflow is thermally driven, whereas hotter flow *always* produces high outflow.

(viii) Outflow could be obtained for both super-Eddington as well as sub-Eddington accretion.

(ix) While constant accretion rate with variable energy solutions prefers a 'weak shock', constant energy with variable accretion rate requires a 'strong shock' to produce high outflow rate.

Although in this work we have performed our calculation for a 10-$M_\odot$ Schwarzschild black hole, general flow characteristics will be unchanged for a black hole of *any* mass except for the fact that the region of parameter space responsible for shock formation will be shifted and the value of $R_{\dot{m}}$ will depend explicitly on the mass of the black hole. In our next work, we will show the direct dependence of $R_{\dot{m}}$ on the mass of the black hole and will apply our model to some specific astrophysical sources of outflows to estimate roughly the amount of outflow in $M_\odot \, yr^{-1}$.

It is to be noted that the primary goal of our present work was to compute the outflow rate and to investigate its dependence on various inflow parameters but *not* to study the collimation procedure of the outflow. In this paper, we have done our calculation in pseudo-Newtonian space–time around a Schwarzschild black hole. Calculation of $R_{\dot{m}}$ from accretion on to a Kerr black hole is in progress and will be presented elsewhere.

## ACKNOWLEDGMENTS

The author would like to thank the unknown referee for providing constructive suggestions.

This paper has been typeset from a T$_{\!E}$X/L$^{\!A}$T$_{\!E}$X file prepared by the author.